# Performance Enhancement of Routers in Networks-on-Chip Using Dynamic Virtual Channels Allocation

Salman Onsori, Farshad Safaei

Faculty of ECE, Shahid Beheshti University G.C., Evin 1983963113, Tehran, IRAN

## ABSTRACT

*This study proposes a new router architecture to improve the performance of dynamic allocation of virtual channels. The proposed router is designed to reduce the hardware complexity and to improve power and area consumption, simultaneously. In the new structure of the proposed router, all of the controlling components have been implemented sequentially inside the allocator router modules. This optimizes communications between the controlling components and eliminates the most of hardware overloads of modular communications. Eliminating additional communications also reduces the hardware complexity. In order to show the validity of the proposed design in real hardware resources, the proposed router has been implemented onto a Field-Programmable Gate Array (FPGA). Since the implementation of a Network-on-Chip (NoC) requires certain amount of area on the chip, the suggested approach is also able to reduce the demand of hardware resources. In this method, the internal memory of the FPGA is used for implementing control units. This memory is faster and can be used with specific patterns. The use of the FPGA memory saves the hardware resources and allows the implementation of NoC based FPGA.*

## 1. INTRODUCTION

With the evolution from single-processor systems to multi-processor systems and the utilization of on chip networks and communication-based systems, many architectural and theoretical studies in various fields have been accomplished. The research fields include design methodology [1-6], topology exploration [5, 7], testing and simulation [8,9], power consumption optimization [10], reliability enhancement [11], and approaches of reducing the area, and latency of the NoC in order to find the most suitable interconnection for application with the SoCs and MP-SoC systems.

The power consumption of communication part of a system must be addressed along with the power consumption of memories and other computing components. Many attempts have been made to reduce the power consumption of communication parts to reduce the overall power consumption of NoCs [12].

Buffers in NoC routers are effective on the overall performance of network [12]. Although other factors can influence power leakage in an NoC, buffers are responsible for about 64% of the total power leakage in an NoC [13]. A buffer also creates dynamic power consumption, which will increase as the current of moving packets in a buffer increases [14]. Furthermore, the area occupied by an on-chip router is dominated by the buffers [15, 16]. Thus, because of the close relationship between the power consumption and the network area and the number of buffers in an NoC, it is necessary to minimize the size of the buffers and also optimize their usage via a proper management without disrupting the efficiency [17].

One effective method for buffer management is the use of virtual channels [18]. Virtual channels were first used to avoid deadlock, but the creation of multiple virtual paths made it possible to avoid head of line (HOL) blocking at the queues, which can also increase the efficiency [19]. There are two methods of static and dynamic management of virtual channels. In the static method, the size and the structure of buffers in a network are determined for the specific traffic pattern. The size is only applicable to a particular application and traffic pattern and cannot change in response to the various traffic patterns in the implementation phase. In the dynamic method, using shared buffer space under a specific traffic pattern, the buffers can be allocated to virtual channels. Providing an accurate design, this method can be very efficient for storing and sending flits using a control unit. Hence, the efficiency of a switch can be maximized by allocating maximum buffer space dynamically. Such single structure with the ability to dynamically allocate buffers is called dynamic allocation of multi-queue (DAMQ) of a buffer. This structure employs a fixed number of queues (virtual channels) for a port. In DAMQ approch, four queues are fortified; three of which are for output ports and one for the interface processor. All packets should be arranged in a queue to be traversed Packets in a queue may fall into the trap because of the first block [20].

DAMQ control unit is complex and its actions are based on a linked list. These linked lists are kept in the pointer registers that must be constantly updated. This causes tricycle latency for the input and output operations of the control units. Most of the latency results from transfer between the pointer registers to update the data. The list of empty buffers should also be updated. Tricycle latency may be acceptable for communications inside a chip, but it is unacceptable for the NoC routers as it causes considerable delay [12].

Other methods are available to simplify the hardware implementation and reduce complexity of the overall DAMQ. DAMQ with self-compacting buffers [21] and a fully-connected circular buffer (FC-CB) [22] are two alternatives based on DAMQ. FC-CB [22] has improved previous schemes [21] by adding a rotational structure that rotates in one direction ensuring that each flit will turn at least one round in each cycle [12].

The usage of FC-CB has two major drawbacks. The first issue arises from complete connections which requires a $P^2 \times P$ crossbar switch for $P$ inputs. The crossbar switch is bigger than the usual square switch and has high power consumption. The second problem is that the rotational shifter allows the incoming flit to be placed anywhere in the buffer; a selective shifter for some parts of the buffer is required so that only required parts will be shifted and the other parts will remain unchanged. This increases the delay, area and power consumption compared to a non-shift method, such as ViCHaR. The resulted overhead is due to the large multiplexers exist between the buffer units for direct and shifted inputs [12].

ViCHaR router has the ability of dynamic allocation of virtual channels based on network traffic. In this method, virtual channel allocation is based on a table, which is a new approach for routers with the ability to dynamically allocate. The method removes the multi-cycle delay created in linked lists via the table. This method presents two new concepts: it employs an integrated buffer structure and each port may have a different number of virtual channels [12]. In this study, the proposed router also employs ViCHaR router methods. The notion of dynamically allocating VC resources based on traffic conditions was presented in [23]. This method has been implemented through VCDAMQ and DAMQ-with-recruit registers (DAMQWR). However, both approaches are coupled with DAMQ underpinnings; hence, they employ the linked-list approach of original DAMQ, which is too costly for an on-chip network [12].

Chaos router [24] and BLAM routing algorithm [25] employ packet misrouting, instead of storage, under heavy load. However, randomized (non-minimal) routing may make it harder to meet strict latency guarantees required in many NoCs (e.g., multimedia SoCs). Further, these schemes do not support dynamic VC allocation to handle fluctuation of traffic.

In [26], a novel dynamic VC architecture is introduced to remove the HOL blockings. In this scheme, the low overhead link list structure is used to manage arriving and departing flits. However, the proposed architecture could not perfectly utilize the unused buffers of their neighboring input channels.

Neishaburi [27] has presented a router for NoC with the ability to evaluate the reliability of virtual channels. This router employs dynamic allocation virtual channels and rational sharing between different input ports. In this way, if a router is being faulty, the virtual associated channels can be collected and allocated to the other input ports. This method does not allow the network bandwidth to be used by the faulty router and exhibits a faulty router to inject the packets; so, it reduces the network traffic, and the average packet delay in the network.

In another study [28], a method for dynamic power management based on virtual channels dynamic allocation has been presented. In this scheme, virtual channels can be dynamically allocated and their number is predicted based on network traffic and the past usage. The number of virtual channels can be predicted based on the history of the past usages of the channels and the present traffic condition of the network. The prediction might increase/decrease the number of virtual channels or keep them the same. A two-stage buffering structure is presented in [29] where the buffers and virtual channels can be dynamically allocated. This approach reduces the latency of the ViCHaR router about 10% ~ 80%, in worst and the best case, respectively.

In [30], a router based on a distributed shared-buffer is suggested which uses some buffers in its output. The buffer control unit in the output has higher efficiency and less delay in queuing. The configuration of the shared buffer structure has lower overhead of power consumption but has also a slight reduction in the performance. Compared with the input buffer router (i.e. the control unit buffers are at the input ports), the router efficiency increases up to 19% and the average packet delay under SPLASH-2 traffic pattern is reduced by 61%. Briefly, it can be concluded that routers with capability of dynamic virtual channels allocation require more complex control units and higher hardware complexity compared to the routers with static allocation. In addition to the hardware complexity, as more hardware units are used, more space is consumed. Introducing a router with dynamic allocation of virtual channels capability can reduce the hardware complexity, power consumption and area usage. Further, the aforesaid capability increases the efficiency and reduces the latency. Moreover, the general purpose architecture is applicable to various traffic patterns. In the proposed router, the control unit and its communications are presented in an optimal fashion. The control mechanism eliminates many additional communications between the control units, which as the result simplifies the hardware design and saves the resources. Hence, the power consumption is improved in comparison to the conventional router.

## 2. MICROARCHITECTURE OF THE PROPOSED ROUTER

The implementation of the proposed router is presented as a pipelined way, which causes the router to produce output in each cycle. The router's control unit is a mixture of combinational and sequential circuits, but the usage of the combinational circuits is negligible. The new control method distributes the control units into basic modules; put it more precisely no separate module is allocated to each unit.

By contrast, in the ViCHaR method, each control unit is implemented separately and as combinational modules to prevent increase in the latency of the router. Sequential implementation of the control units and the use of a clock cycle significantly increases the latency because a time cycle is needed for each storage operation in a control register. Since storage in a router control unit is so common, in the case that the architecture is not suitable, it can significantly increase the number of cycles and consequently cause considerable increase in latency.

The proposed router uses a new method for implementation of the control unit. Control units are implemented sequentially to prevent an increase in the critical path latency. Based on the hardware implementation of the switch and virtual channel allocators, and amount of hardware delays in aforementioned modules (due to the buffers in the pipeline stages), the control units can be implemented inside them. The control units can use the information in the allocator in a pipeline manner, which will then remove the overhead of recreating the information in any of the separated combinational units, which in turn saves the hardware resources. This means that, in a particular cycle of the allocator, controlling actions are implemented and operate in parallel to other sections of the module that are independent of them. Likewise, in the design phase, all control sections in the router modules, especially virtual channel and switch allocator, are implemented in parallel. There is no longer a need for large combinational circuits to implement the control units in parallel with different components of the router. This reduces the number of router registers and surface area consumption.

The proposed router presents a new architecture in which additional controlling communications have been removed and optimized. The control components have been restructured to optimize their communications. This reduces the hardware complexity by reducing the hardware resources, improving controlling communications, and by the dynamic allocation of virtual channels, which is a major challenge in the design of such routers.

In order to understand the structure of the proposed router, many factors such as the number of input and output ports, the type of control unit, and the mode of implementation should be considered. The proposed router has five input and output ports through which input and output of flits takes place. Four ports are connected to adjacent routers and one remaining port is connected to the computing section of the router. The computing section can be memory, a processor, a sound or image processor, or some other computational element. In the implementation, packets with size of four flits have been used. A control unit consists of a header flit, two body flits and a tail flit. Each flit is 128 bits. Bits 0 and 1 of each flit specify the type of flit. Figure 1 shows the structure of a flit and its section types.

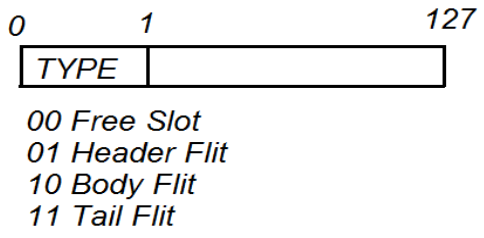

Figure 1. The format of Flit and meaning of TYPE field

Figure 1 reveals that in a unified buffer structure (UBS), the first two bits of each flit can have one of four values. If the value is "01", it is a header flit; if it is "10", then it is a body flit; and if it is "11", it is a tail flit. The "00" state represents an empty buffer. Each arriving flit is checked to determine its type. If it is a header flit, it is sent in parallel to the routing unit (to calculate its output port according to the destination) and UBS (for storage). The selection scheme of the buffer for this flit is based on the storage unit of available buffers. The unit notices which of the first two bits of the buffer in UBS is empty. If the first two bits are "00", the buffer is empty and it is allocated to the flit. If any other flits than the header enter the router, it will go through the same process for allocation in the buffer. If the entering flit is a body or tail flit, after placement in UBS, its header flit should be checked. If it is specified from the control table that a virtual channel is allocated to its header flit, that special virtual channel will be allocated to the flit and also the number of this flit will be written to the control table; therefore, the control table will be updated. If it is not a body or tail flit, it will remain in UBS and wait for virtual channel allocation as a header.

For each input port in the router, UBS contains 16 buffers. This structure is integrated and resembles the structure of a buffer in a virtual channel wormhole router except that the control structure is integrated. Since the router has five input ports and each port has 16 buffers, the virtual channel allocation comprises twenty five 16:1 arbiters. In addition to 16:1 arbiters of arbitration of the first stage, five 5:1 arbiters are needed for the second stage. Arbiters are sequential and arbitration is based on constant priority. Each sequential arbitration consists of two arbiters based on simple priority with using of mask.

The switch allocator is composed of two judging stages. In the first stage, five 16:1 arbiters are used and, the second phase uses five 5:1 arbiters. In the first stage, the arbitration is made between flits of the virtual channel pointed with the flit output pointer. For each virtual channel, a request can exist, which is why 16:1 arbiters are needed. After selection by the switch allocator, a signal is sent to UBS to read the buffer and transfer the flit to the crossbar switch. The table must also be updated and the corresponding output port should be transferred to the next location.

The proposed router has been implemented on a FPGA to provide an accurate hardware sample of the proposed router that is more realistic than software simulation. It was implemented and synthesized using ISE 12.3 software for programmable gate arrays (Xilinx). The type of programmable gate array should be determined for using its available resources to synthesize the design. Since the programmable gate arrays are limited in hardware and each group can implement a hardware design of a specific complexity, Virtex 6 was used for synthesis of the router. The chip is very advanced and very powerful in terms of computing capabilities.

Using the sequential method does not increase the latency of the router compared to ViCHaR because the sequential control components in this method vary by different modules and according to the relationship between them and router timing. To keep the latency fixed, the modules record the control units in registers as needed (for timing purposes). This can be done with the knowledge of the latency of the router's modules and their relationships will prevent an increase in latency. Figure 2 shows the relationship between the virtual channel and switch allocators in the proposed router.

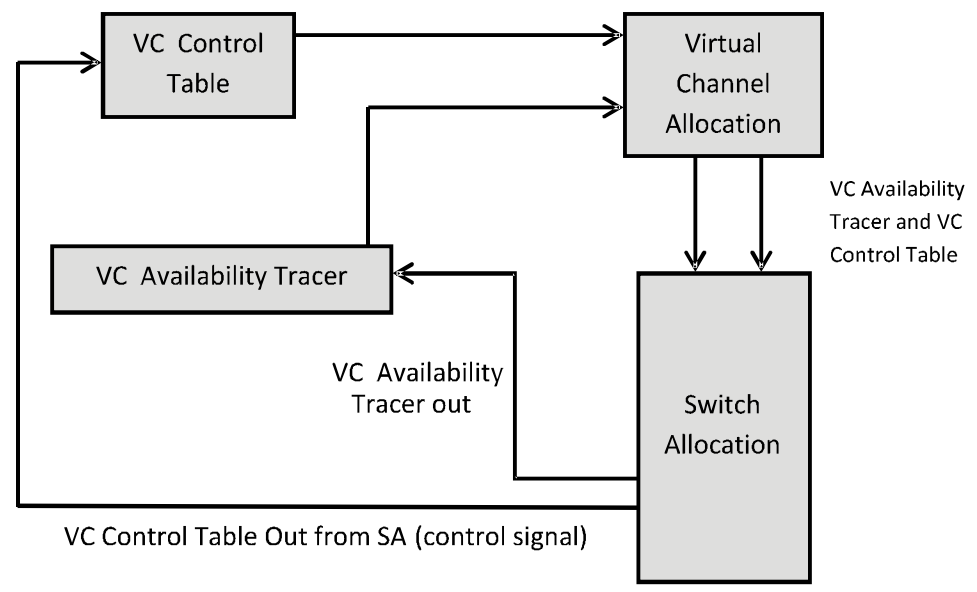

Figure 2. The relationship between VC allocator and switch allocator in the proposed router

In this figure, virtual channel allocator holds virtual channel control table and virtual channel availability tracer and uses these units for allocating virtual channels. Virtual channel allocator will update virtual channel control table and virtual channel availability tracer after allocation. The update process occurs in virtual channel allocator because it has many of signals which are necessary for updating control table and we don't need to send these signals to the output ports to be used in combinational control logics. This approach makes the hardware design simpler. Virtual channel control table is updated and can be used by switch allocator at the final stage of the module operation. After that, virtual channel allocation sends these two control signals to switch allocator as these control units were updated in virtual channel and they contain the accurate values. Therefore, two allocators continue their operation with correct data and control units work efficiently and also the router delay does not increase. Token dispenser is designed in virtual channel allocator unit and operates with the router clock. It uses virtual channel control table and virtual channel availability tracer data to allocate virtual channels.

In ViCHaR, the virtual channel allocator transfers information to its outputs to be used to control components communicate with each. Hence, the new virtual channel will be allocated. This relationship is demonstrated in Figure 3.

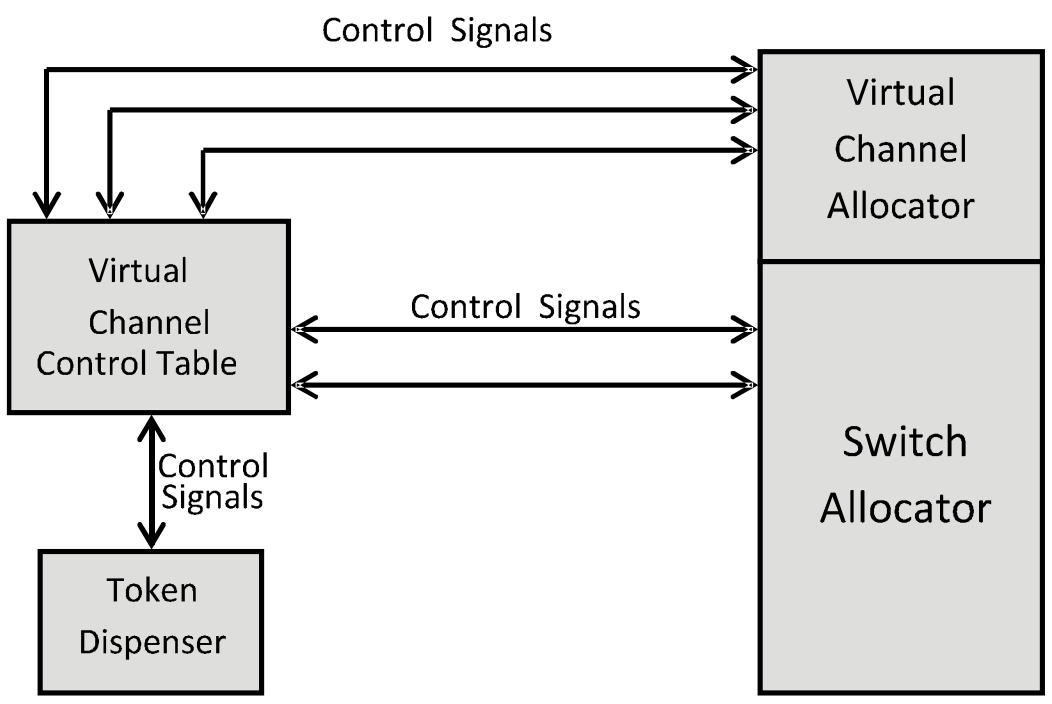

Figure 3. Control connection between switch and VC allocator in ViCHaR router

As shown in Figure 3, allocators receive the control signals from virtual channel control table, which is implemented in a combinational way. At the end of the operation, allocators send out signals which contain new results to VC control table. The combinational section of the control table is responsible for updating the table and communicating with the other control parts like token dispenser.

In the proposed router, by contrast, all the control functions in switch and virtual channel allocation unit are performed sequentially which reduces both the hardware complexity and the communication of control units. Consequently, there would be no need to continuous communication of the allocators with external control units. This integration in control components optimizes the usage of hardware resources in the field programmable gate arrays and reduces the power consumption by using fewer registers.

Figure 4 depicts the structure of virtual channel allocator in the proposed router. As shown in the figure, token dispenser is embedded in the virtual channel allocation unit which allocates virtual channels based on the information of virtual channel control table and virtual channel availability tracer units along each other.

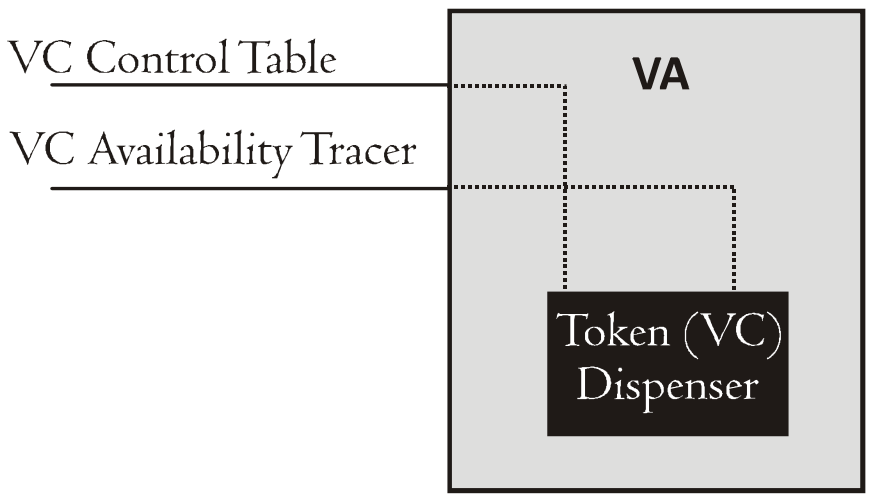

Figure 4. The structure of virtual channel allocator in the proposed router

Slot availability tracer is updated in combination with UBS. When a flit enters UBS or leaves a crossbar switch, the unit is updated by UBS. Slot availability tracer is also sent to neighboring routers as a credit. In Figure 5, the relationship between ~~the~~ UBS and slot availability tracer is illustrated.

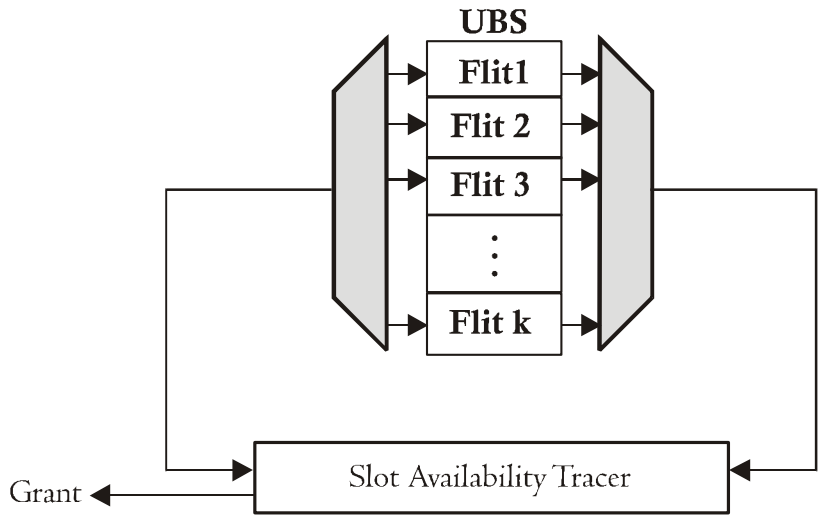

Figure 5. Interfacing UBS and Slot availability tracer for one input

The overall design of the proposed router and the relationships between its components are depicted in Figure 6. The buffering structure of UBS is illustrated for just one input port. The router is based on ViCHaR router and it can dynamically allocate virtual channels using a table. The control components of the router are implemented sequentially. The router uses buffering technique for control units and by using them in suitable times according to communications between different components of the router; control units are implemented in different modules and they are not in separate and combinational form. As a result, communication between different modules and control units becomes simpler, and the hardware resources for implementation of the router on programmable gate array and the complexity of the control unit will be decreased.

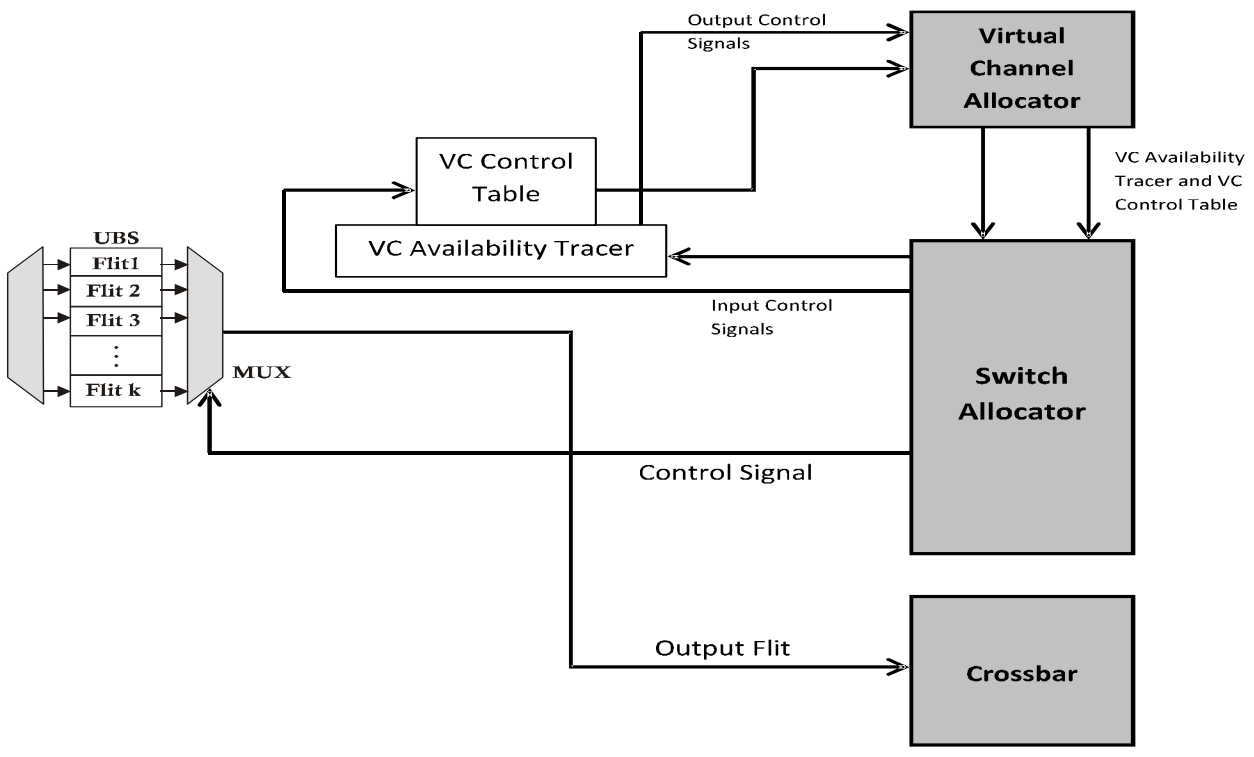

Figure 6. The architecture of the proposed router

Programmable gate arrays have multiple sources for different applications, which improve the implemented design. Some properties need more sophisticated implementation to be introduced for the synthesis tool to be able to identify these resources. In this study, a new approach for the implementation of the router is presented that makes efficient usage of hardware and software implementation. Moreover, it reduces area and power consumption in programmable gate arrays.

The control table has been implemented to be embedded with arriving/departing flit pointers. The table has the ability to integrate control of the switch and virtual channel allocators. Some programmable gate arrays have internal memories which can be used in different implementations. Vertix6 has this property. The memory is used appropriately to reduce the delay caused by hardware and software and decrease the power and area consumption. Virtex6 has a specific and fixed number of internal memories which are faster. Also, it has better performance than registers and programmable parts. The usage of mentioned memories decreases the number of registers and programmable parts, so it increases the speed and performance of the router.

The control table in the new implementation will be in embedded memories of programmable gateway array. Since the memory embedded in programmable gateway array has less speed and power consumption, there will be a decrease in the power consumption.

To employ the embedded memories in FPGA, a specific pattern expressed in the documentation of Xilinix ISE software must be followed. The special modules should be defined to force the synthesis program to use embedded memories instead of registers and programmable parts for implementing hardware designs in programmable gate array. Modules in embedded memory (control table) operate concurrently with other control units of the router. If control units of the router were not implemented sequentially in virtual channel and switch allocators, the implementation of control table into embedded memory would not be possible because the memories should be defined sequentially to be synthesized with the program. For example, using this method is impossible for control units of ViCHaR router because they are combinational logics.

The implementation of control table in internal memories of programmable gate arrays decreases the usage of hardware resources; hence, it would be possible to implement a complete NoC structure on a FPGA chip. Moreover, this implementation reduces power consumption of the router.

In what follows, the results captured by simulations will be discussed in details. The experimental results for the proposed router and ViCHaR router on the gate arrays were obtained using Xilinx ISE Design Suite 12.3. This software package calculates the power and area consumption of implemented hardware designs on Xillinx programmable gate arrays.

In Figure 7, the percentage of the total delay for each modules of the proposed router is reported. Based on the figure, virtual channel allocator (VA) has the lowest frequency and the highest delay.

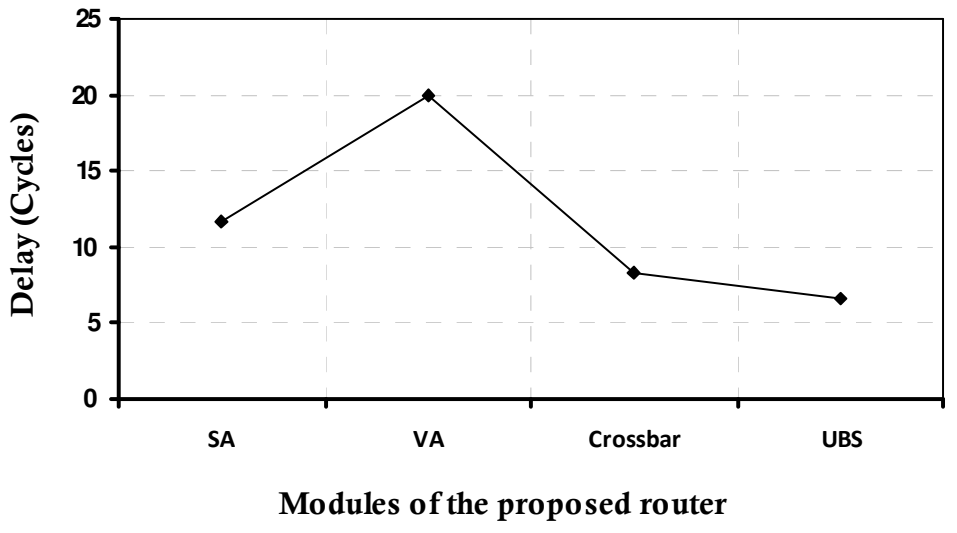

Figure 7. Percentage of the total delay for each module in the proposed router

As shown in Figure 8, the virtual channel allocator has the lowest operating frequency. Since this module is responsible for the implementation of the control processes, the more hardware was used and the higher level of delay was experienced. As a result, the operating frequency decreases.

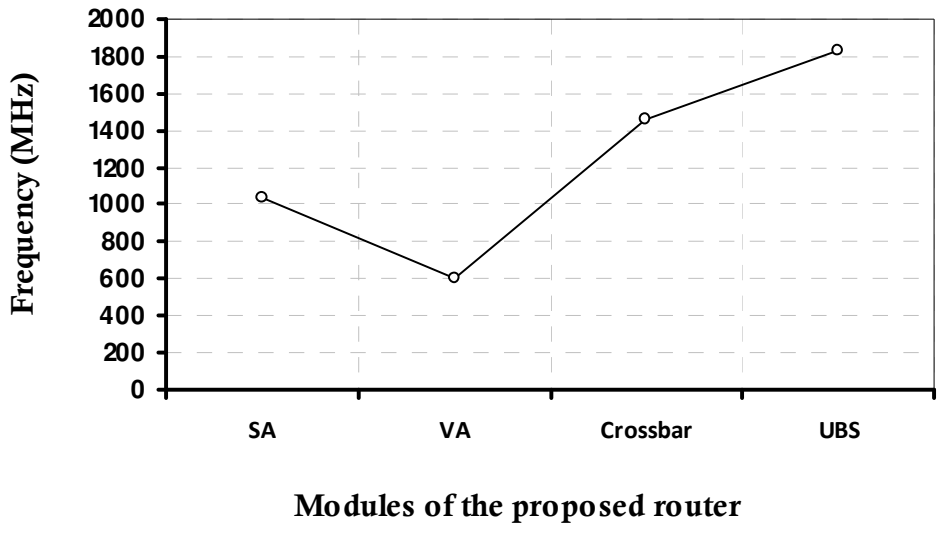

Figure 8. Modules' frequency in the proposed router

In Figure 9, the dynamic power consumed by the router is shown for comparison. There is much enhancement in the ViCHaR router; the dynamic power consumption of ViCHaR router after implementation on FPGA is demonstrated in this figure.

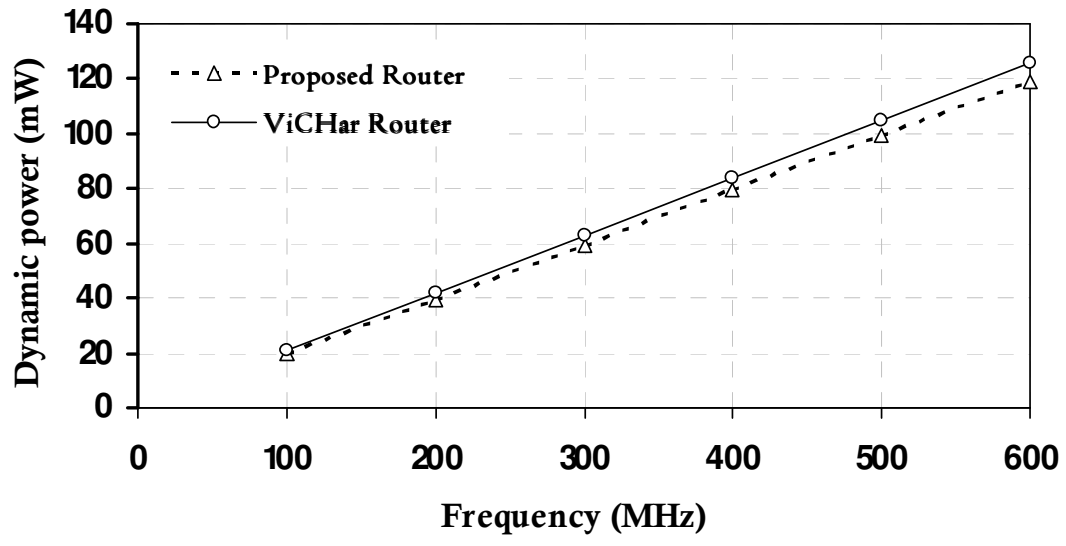

Figure 9. Comparison of dynamic power in the proposed router and ViCHaR router

The dynamic power consumption in the proposed router diminished in comparison to the ViCHaR router. Figure 10 shows this improvement in terms of dynamic power consumption in more details. The improvement of power consumption of the router at different frequencies over the ViCHaR routers after implementation on FPGA is illustrated in this figure.

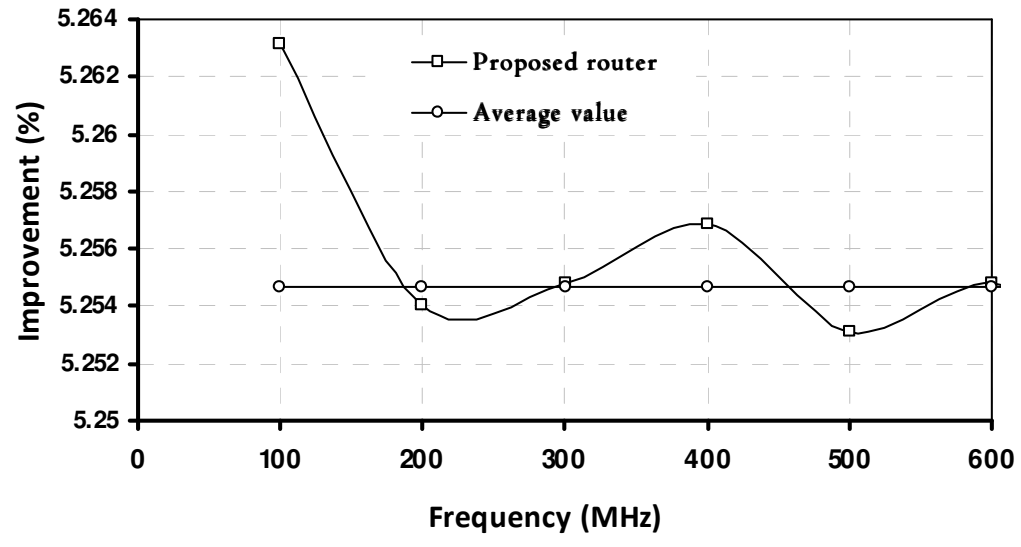

Figure 10. Percentage of the Improvement in the proposed router

As can be seen in Figure 10, there was about 5% improvement in power consumption of the proposed routers. This value differed slightly at different frequencies. The line in the figure delineates improved frequencies at about 5.25%. As expected, removing additional communication between control units and utilization of the unified control structure lead to reduction of power consumption.

In Figure 11, the hardware used in the proposed router and the ViCHaR router are shown. These results were synthesized and implemented with Xilinx synthesis programs. The figure indicates that proposed router reduces the need to hardware resources. According to the figure, it is clear that the area consumption in the proposed router decreases compared to the ViCHaR router. Figure 11 compares three types of the hardware utilized in FPGA. These resources are compared with each other and marked in the figure.

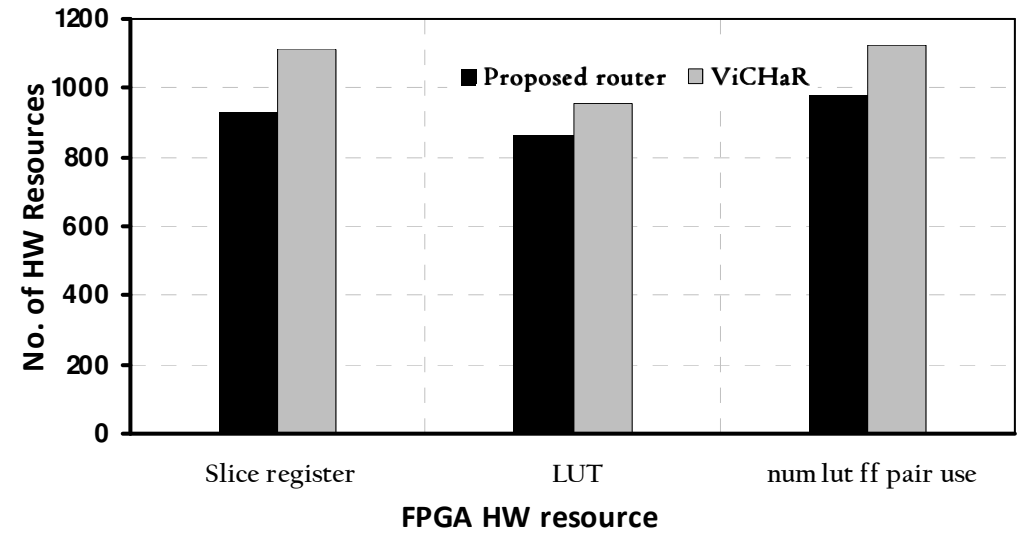

Figure 11. Comparing the number of FPGA hardware resources in the proposed router vs. ViCHaR

Figure 12 better illustrates the hardware improvement and the reduction of the area in the programmable gate arrays after implementing the proposed router. The percentage of improvement for each component along the general improvement in area is provided in the figure.

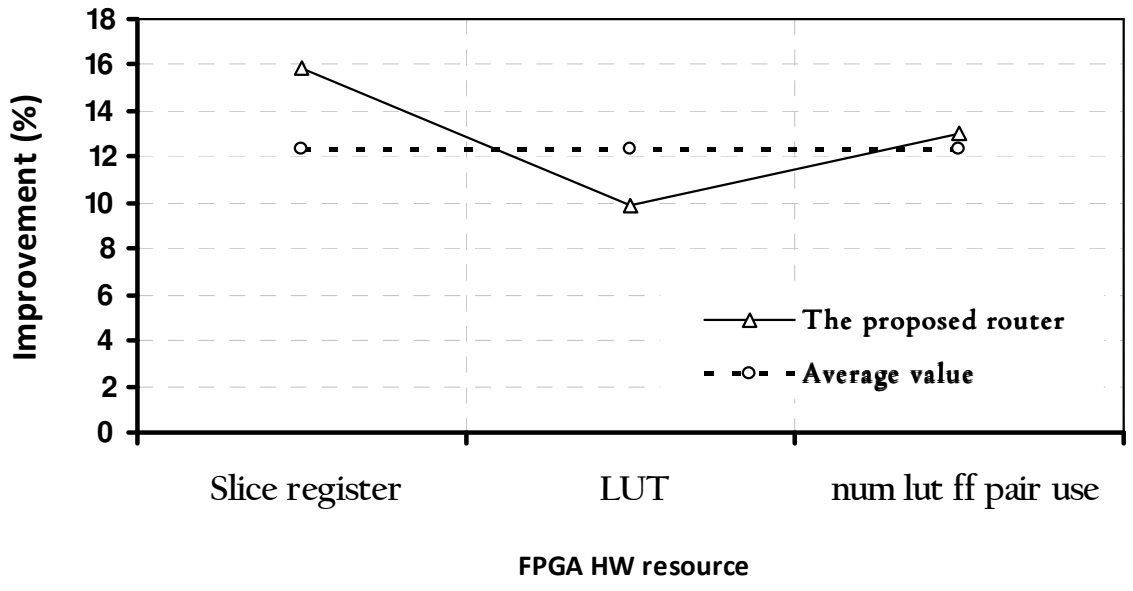

Figure 12. Percentage improvement of hardware resources in the proposed router

As shown in the figure, each hardware resource is improved by some percentage. Improvements in the hardware and the reduction in area are shown as a solid line delineating the improvement in the weighted average. General improvements in the area and power consumption were about 12.3% which was expected since most of communication-related overhead in the new router was eliminated. The power consumption for logical units and buffers is plotted in Figure 13.

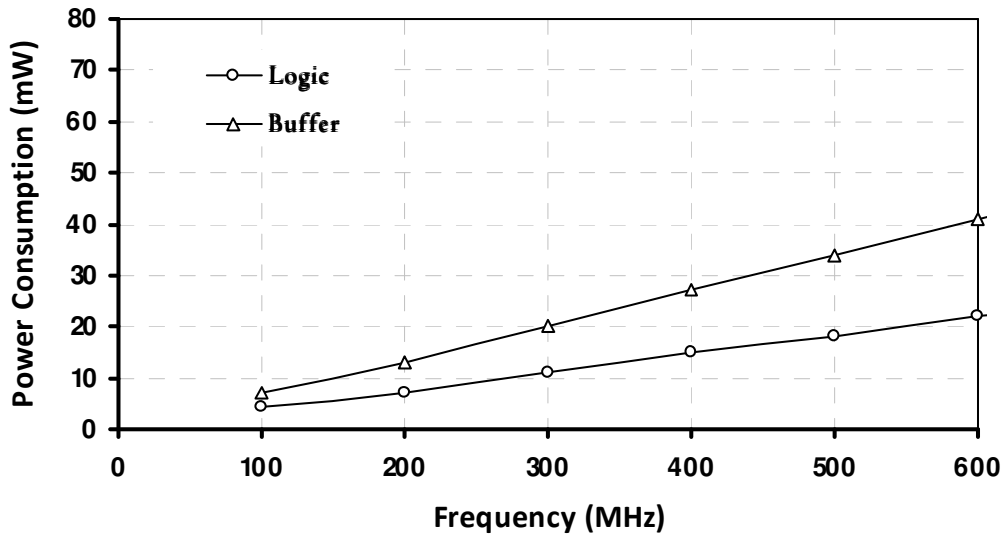

Figure 13. Power consumption of the buffers and logics in the proposed router

According to Figure 13, the power consumption of the proposed router's buffers is about twice of the implemented logical units. This figure emphasizes the importance of buffer management schemes in NoC. Power consumption of a router highly depends on its buffers which also is clear from power consumption in the proposed router. At the different frequencies, the ratio of power consumption of router's buffers to the total power consumed by the router is shown in Figure 14.

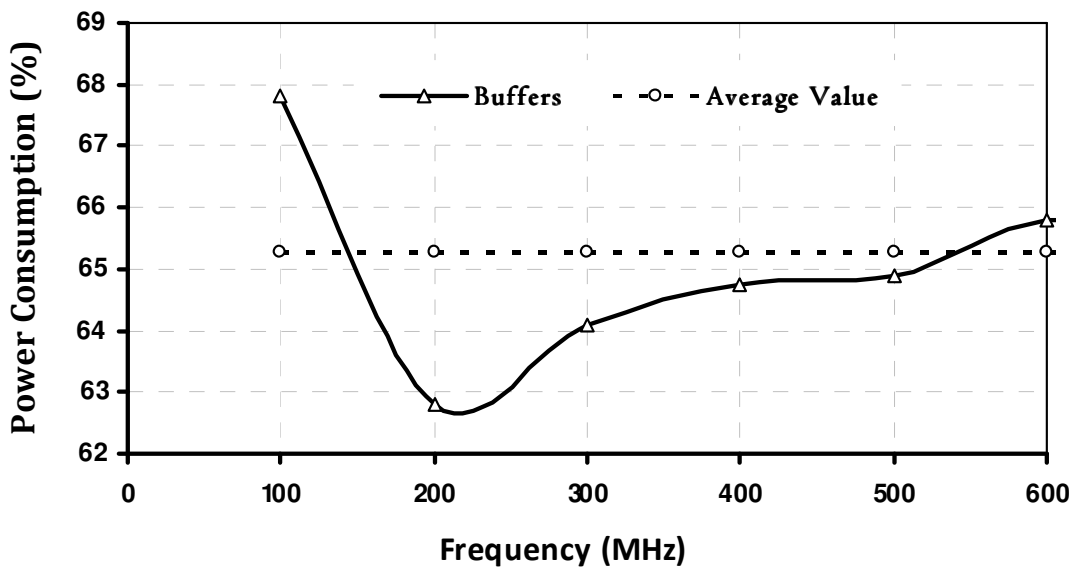

Figure 14. Percentage of buffer's power consumption from the total power of the proposed router

According to the figure, the consumed power of buffers constitutes 65% of the total power consumption in the presented router; this amount is also delineated linearly, which shows the average amount at different frequencies. This result was reported in the previous work and the results of the proposed router also confirm it.

In Figure 15, the router leakage power is shown at different frequencies. The amount of leakage power is far more than other consumed powers and it will increase by frequency.

As explained before, the proposed router is implemented in two ways. The second method of was done after the first stage and improves the results of the first implementation. At the first implementation, the idea of unified and optimized control components was implemented and results are presented in the figure. According to the results, the suggested router improves the hardware resources, power, and area consumption. To implement the second router, a new approach for implementation and usage of hardware resources was employed. We used the internal memories of programmable gate arrays to implement the control table in the second method which increases the efficiency of the first implementation.

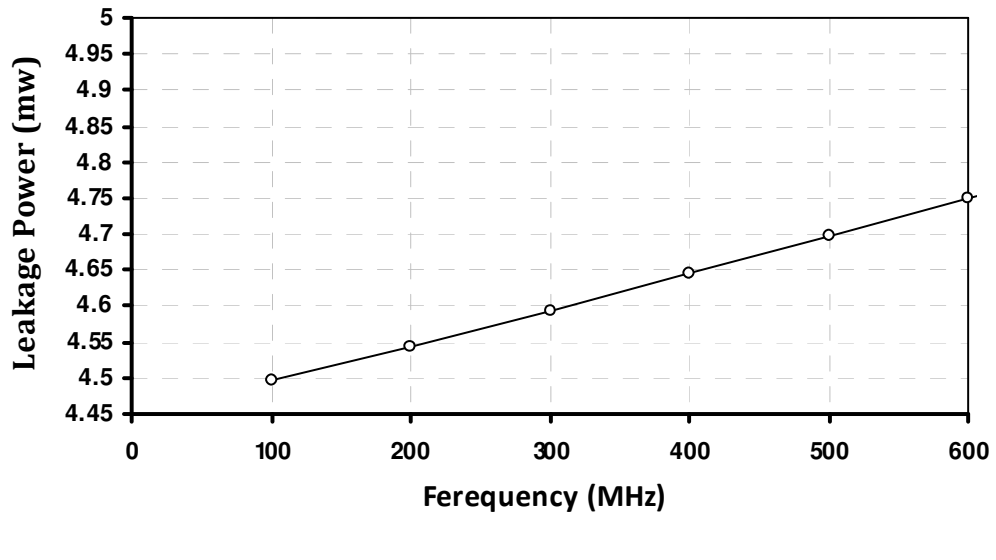

Figure 15. The leakage power of the proposed router

In Figure 16, the consumed power of the second implementation is presented for 16, 32, 64, and 128-bit flits with an operating frequency of 500 MHz.

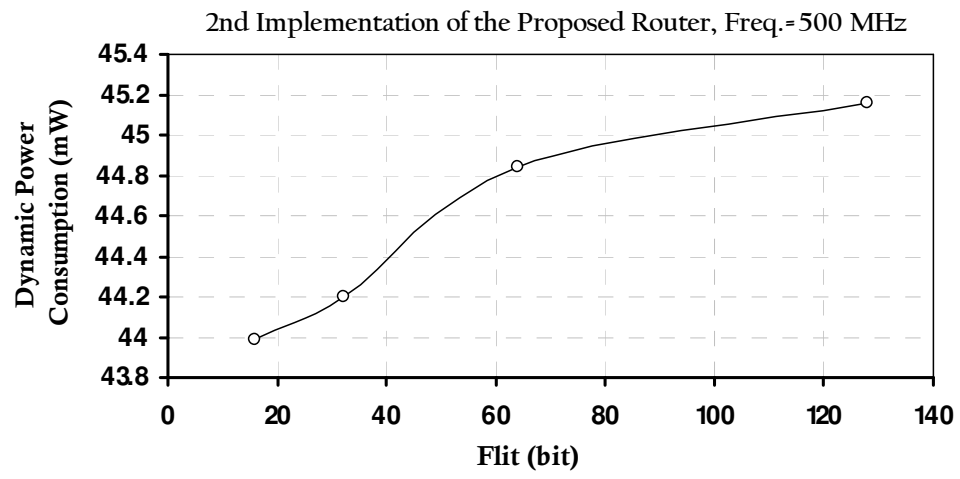

Figure 16. Dynamic power consumption of the second implementation in terms of flits with operating frequency of 500 MHz

In this figure, the implemented router power consumption with the second approach is presented at different frequencies. In addition, power consumption of the first implementation of the router is presented for comparison.

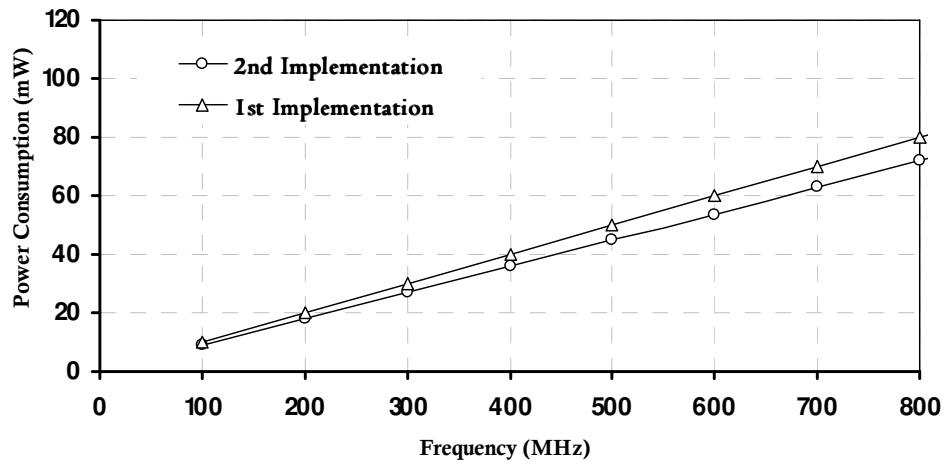

Figure 17. Comparing the power consumption of the 1st and 2nd implementations of the proposed router

For a better understanding of the improvements made at different frequencies, Figure 18 is plotted. In this figure, the percentage of power reduction using the second implementation of the router instead of the first one is shown against with different frequencies.

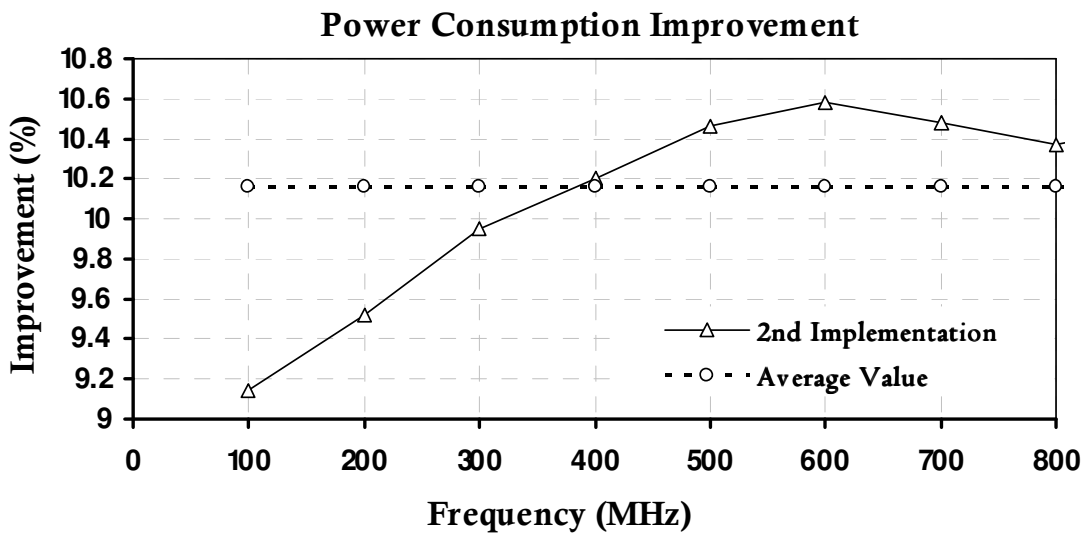

Figure 18. Percentage of the power consumption improvement in the 2nd implementation of the proposed router using embedded memories

As shown in the figure, the mean value of power reduction is about 10.15%. It means that the implementing of the control table using the embedded memories in programmable gateway arrays will reduce the power by 10%.

## 3. CONCLUSIONS

In this study, a new router with the ability of dynamic allocation of virtual channels is presented. The router allocates virtual channels based on a table. In the allocation method provided by ViCHaR, the control unit is complex. The proposed router provides a new control unit that eliminates many unnecessary controlling communications, saves the hardware resources of the control units, and considerably reduces the hardware complexity. The router provides a new architecture in which the control components are designed in a new way. It improves the control structure and reduces the power consumption of the ViCHa Rrouter. To show the improvements made in the proposed router, it was implemented on programmable gate arrays and the results of the simulation were compared to the results obtained from the ViCHaR router. The experimental results suggest a reduction in power consumption, area consumption, and complexity of the proposed router compared to the ViCHaR router.

## References

[1] W.J. Dally and B. Towles, “Route packets, not wires: On-chip interconnection networks,” in Proc. Des. Autom. Conf., Jun. 2001, pp. 684–689.

[2] L. Benini et al., “Networks on chips: A new SoC paradigm,” in IEEE Computer, vol. 36, no. 1, pp. 70–78, Jan. 2002.

[3] S. Kumar, "A network on chip architecture and design methodology," in Proceedings of the IEEE Computer Society Annual Symposium on VLSI (ISVLSI.02), pp. 105-112, 2002

[4] K. Goossens, J. Dielissen, A. Radulescu, “The Æthereal network on chip: Concepts, architectures, and implementations,” IEEE Design and Test of Computers, vol. 22, no. 5, pp. 414–421, Sept.-Oct. 2005.

[5] D. Bertozzi and L. Benini, “Xpipes: a network-onchip architecture for gigascale systems-on-chip,” IEEE Circuits and Systems Magazine, vol. 4, no. 2, pp. 1101-1107, 2004.

[6] M. B. Stensgaard and J. Sparsø, "ReNoC: A Network-on-Chip Architecture with Reconfigurable Topology," Second ACM/IEEE International Symposium on Networks-on-Chip, pp. 55-64, 2008.

[7] P. P. Pande et al., "Performance Evaluation and Design Trade-Offs for Network-on-Chip Interconnect Architectures," IEEE Transaction on Computer, vol. 54, no. 8, pp. 1025- 1040, 2005.

[8] N. Banerjee et al., "A Power and Performance Model for Network-on-Chip Architectures," In Proceedings of the Design, Automation and Test in Europe Conference and Exhibition (DATE’04), vol. 2, pp. 1250- 1255, 2004.

[9] H-S Wang, L-S Peh, S. Malik, "Orion: A Power- Performance Simulator for Interconnection Network," in International Symposium on Micro architecture, Istanbul, Turkey, pp. 294-305, November 2002.

[10] K. lee et al., "Low-Power Network-on-Chip for High-Performance SoC Design," IEEE Transactions on Very Large Scale Integration (VlSI) Systems, vol. 14, no. 2, pp. 148-160, 2006.

[11] D. Park et al., " Exploring fault-tolerant network-on-chip architectures," in Proceedings of the 2006 International Conference on Dependable Systems and Networks (DSN’06), pp. 93-104, 2006.

[12] C. A. Nicopoulos , D. Park, J. Kim, N. Vijaykrishnan , M. S. Yusef, C. R. Das, "ViCHaR: A Dynamic Virtual Channel Regulator for Network-on-Chip Routers", in International Symposium on Microarchitecture, pp. 333-346, 2006.

[13] C. Xuning and L. S. Peh, "Leakage power modeling and optimization in interconnection networks," in Proceedings of the International Symposium on Low Power Electronics and Design (ISLPED), pp. 90-95, 2003.

[14] T. T. Ye, L. Benini, and G. De Micheli, "Analysis of power consumption on switch fabrics in network routers," in Proceedings of the 39th Design Automation Conference (DAC), pp. 524-529, 2002.

[15] G. Varatkar and R. Marculescu, "Traffic analysis for on-chip networks design of multimedia applications," in Proceedings of the 39th Design Automation Conference (DAC), pp. 795-800, 2002.

[16] H. Jingcao and R. Marculescu, "Application-specific buffer space allocation for networks-on-chip router design," in Proceedings of the IEEE/ACM International Conference on Computer Aided Design (ICCAD), pp. 354-361, 2004.

[17] J. D. Owens, W. J. Dally, R. Ho , D. N. Jayasimha, S. W. keckler, L. S. Peh , "Research Challenges for On-Chip Interconnection Networks," in IEEE Micro, vol. 27, 2007.

[18] W. J. Dally, "Virtual-channel flow control," in Proceedings of the 17th Annual International Symposium on Computer Architecture (ISCA), pp. 60-68, 1990.

[19] C. A. Nicopoulos, Network-on-Chip Architectures: A Holistic Design Exploration, Ph.D. Thesis, Department of Electrical Engineering, Pennsylvania State University, 2007.

[20] Y. Tamir and G. L. Frazier, "High-performance multi-queue buffers for VLSI communications switches", SIGARCH Comput. Archit. News, vol. 16, pp. 343-354, 1988.

[21] J. Park and B. W. O'Kraftam, "Design and evaluation of a DAMQ multiprocessor network with self-compacting buffers", in Proceedings of Supercomputing, pp. 713-722, 1994.

[22] N. Ni, M. Pirvu, L. Bhuyan, "Circular buffered switch design with wormhole routing an virtual channels" in ICCD ’98: Proceedings of the International Conference on Computer Design, pp. 466-473, Oct. 1998.
[23] Y. Choi and T. M. Pinkston, "Evaluation of queue designs for true fully adaptive routers," in Journal of Parallel and Distributed Computing, vol. 64, no. 5, pp. 606-616, 2004.
[24] S. Konstantinidou and L. Snyder, "The Chaos router," IEEE Transactions on Computers, vol. 43, pp. 1386-1397, 1994.
[25] M. Thottethodi, A. R. Lebeck, S. S. Mukherjee, "BLAM: a high-performance routing algorithm for virtual cut-through networks," in Proceedings of the International Parallel and Distributed Processing Symposium (IPDPS), pp. 10 pp., 2003.
[26] M. Lai, Z. Wang, L. Gao, H. Lu, K. Dai, "A Dynamically-Allocated Virtual Channel Architecture with Congestion Awareness for On-Chip Routers," in Proceedings of the 46th Design Automation Conference (DAC), pp. 630-633, 2008.
[27] M. H. Neishaburi, and Z. Zilic, "Reliability Awar NoC Router Architecture Using Input Channel Buffer Sharing", in ACM Great Lake Symposium on VLSI(GLSVLS), pp. 511-516, 2009.
[28] A. M. Rahmani, and Z. Zilic, "Forecasting-based Dynamic Virtual Channels Allocation for Power Optimization of Network-on-Chips", in International Conference of VLSI Design, pp. 151-156, 2009.
[29] C. Concatto, , A. Kologeski , L. Carro, and F. Kastensmidt, "Two-Levels of Adaptive Buffer for Virtual ChannelRouter in NoCs", in International Conference of VLSI and System on Chip, pp. 302-307, 2011.
[30] R.S. Ramanujam, V. Soteriou, B. Lill, L. S. Peh, "Extending the Effective Throughput of NoCs with Distributed Shared-Buffer Routers", IEEE Transactions on Computer-Aided Design of Integrated Circuits and Systems, vol. 30, no. 4, 2011.

## Authors

**Salman Onsori** He received the B.S. degree from Shahed University, Tehran, Iran, in 2010, and the M.S. degree from Shahid Beheshti University (SBU), Tehran, Iran, in 2013, both in computer engineering. His current research interests include Networks-on-Chip architecture, multiprocessor Systems-on-Chip, and Embedded systems.

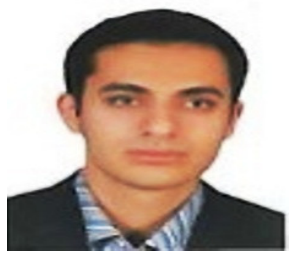

**Farshad Safaei** He received the B.Sc., M.Sc., and Ph.D. degrees in Computer Engineering from Iran University of Science and Technology (IUST) in 1994, 1997 and 2007, respectively. He is currently an assistant professor in the Department of Electrical and Computer Engineering, Shahid Beheshti University, Tehran, Iran. His research interests are performance modelling/evaluation, Interconnection networks, computer networks, and high performance computer architecture.

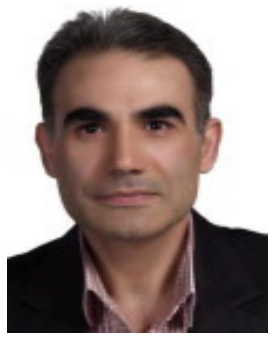